\documentclass[prl,twocolumn,superscriptaddress]{revtex4}

\def\cm-1{cm$^{-1}$}

\begin{document}

\textbf{Blumberg \textit{et al.} Reply:} 
The recent scanning SQUID \cite{Tsuei}, tunneling 
\cite{Biswas}, ARPES \cite{Sato,Armitage1}, penetration 
depth \cite{Kokales,Prozorov,Skinta} and specific heat 
\cite{Balci} experiments provided convincing 
evidence of $d_{x^{2}-y^{2}}$-wave pairing symmetry for the 
electron-doped cuprates at low- and optimal doping. 
In agreement with these experiments the recent electronic Raman 
scattering studies \cite{blumberg02} rule out an anisotropic 
$s$-wave scenario: in contrast to conventional $s$-wave superconductors 
no gap-threshold structure has been observed in electronic Raman response 
for any symmetry channel even at the lowest temperatures and 
frequencies measured. 
On the other hand, the pair breaking excitations measured by 
polarized electronic Raman scattering indicate a larger magnitude of 
the superconducting (SC) gap closer to the middle of the Brillouin zone 
(BZ) quarters, the vicinity of $(\pm \pi/2a, \pm \pi/2a)$ points, than 
to the BZ boundaries \cite{blumberg02,stadlober95}.
The latter results are inconsistent with the monotonic  
$d_{x^{2}-y^{2}}$ SC order parameter (OP) function, 
$\Delta({\bf{k}}) \propto \cos({k_{x}a}) - \cos({k_{y}a)}$,  
where $\bf{k}$ is a wave vector on the Fermi surface (FS) and $a$ is 
the $ab$-plane lattice constant. 

Our proposal of a nonmonotonic $d_{x^2-y^2}$ OP for which the positions 
of the SC gap maxima are located closer to the nodes than to the BZ 
boundaries \cite{blumberg02} reconciles all experimental 
observations \cite{note1}. 
Indeed, the maximum Raman gap, $2\Delta_{B_{2g}} = 67$~\cm-1, is 
consistent with the gap value of $\Delta^{tunn}_{max} = 3.7$~meV observed 
in tunneling spectroscopy \cite{Huang:90} and 
the $2\Delta_{B_{1g}} = 50$~\cm-1 is 
consistent with the leading edge gap at the BZ boundary  
estimated from ARPES experiments \cite{Sato,Armitage1}. 
For hole doped cuprates, superconductors with short correlation 
length, the nearest neighbor correlation is strongest and monotonic 
OP is expected. 
In contrast, for electron doped cuprates 
the superconducting correlation length is long (low upper critical 
fields) \cite{Wang} indicating the importance of further correlations  
leading to a nonmonotonic OP. 

In the preceding Comment Venturini \textit{et al.} \cite{Venturini} 
note that for a nonmonotonic OP a multiple peak/shoulder structure 
is expected in the Raman response. 
Their calculation (Fig.~1a), however, does not account for realistic 
FS topology, energy and momentum dependent relaxational behavior, 
possible impurity scattering rates and inhomogeneous broadening, and 
is sensitive to the gap functional form. 
Unrealistically small constant phenomenological damping $\Gamma = 
1.3$~\cm-1 has been used, while, for example, the best fit for the 
Bi$_{2}$Sr$_{2}$CaCu$_{2}$O$_{8}$ compound required $\Gamma = 
43$~\cm-1 \cite{Devereaux}, much larger than the separation in fine 
structure of DOS for our proposed nonmonotonic OP, 
$\Delta_{B_{2g}} - \Delta_{B_{1g}} = 8.5$~\cm-1. 
For larger $\Gamma$ the sharp singularities are not expected to 
be resolved and indeed   
a flat top structure for spectra in the $B_{1g}$ channel 
is experimentally observed (Fig.~3 in \cite{blumberg02}). 
Calculations with larger phenomenological damping (Fig.~1b in 
\cite{Venturini}) well resemble experimental data at high frequencies,  
however, the use of momentum and energy independent damping for in-gap 
energies is not justified even for strongly 
anisotropic superconductors \cite{Coffey,Kaminski}: the  
imposed large values of $\Gamma/\omega$ are unphysical and 
wipe out characteristic low-frequency power laws. 

The low-frequency power laws in Raman response indeed provide 
independent verification 
for existence and position of the SC gap nodes. 
The experimentally observed power laws, approximately cubic for $B_{1g}$ 
and linear for 
$B_{2g}$ channels (Fig.~3 in \cite{blumberg02}), are in agreement with 
the expectations for $d_{x^{2}-y^{2}}$-wave superconductors: 
corresponding power laws rise from the convolution of the constant density 
of states in the vicinity of the nodal Dirac points and the Raman 
scattering amplitude squares, $\gamma^{2}_{B_{1g},B_{2g}}(\phi)$ 
\cite{Venturini,Devereaux,note2}. 

We note that for a nonmonotonic OP the phase 
space of the nodal regions is reduced relative to the monotonic OP.     
As a result, a stronger activated-like contribution in thermodynamic 
properties is expected. 
Measurements down to sub-Kelvin temperatures are required to 
emphasize the nodal behavior:  
the power laws are dominant only below a 
crossover temperature that depends on the nodal velocity.
The disagreements of thermodynamic and penetration depth measurements 
at relatively high temperatures 
with fits to monotonic $d$-wave OP are therefore not surprising. 

\vspace{1mm}

G.~Blumberg, A.~Koitzsch, A.~Gozar, and B.S.~Dennis, 
Bell Laboratories, Lucent Technologies, Murray Hill, NJ 07974 

C.A.~Kendziora, US Naval Research Laboratory, Code 6375, Washington, 
D.C. 20375 

P.~Fournier and R.L.~Greene, 
University of Maryland, College Park, MD 20742


\begin{thebibliography}{99}
\bibitem{note1} 
Note that the nonmonotonic function $2\Delta(\phi)$ shown in 
\cite{blumberg02}, Fig.~1B, is a sketch. 
Raman measurements do not resolve the precise gap maximum angle 
$\phi_{max}$.  
Moreover, the nodal velocity and the $\phi_{max}$ are expected to be a 
function of doping concentration.
\bibitem{note2} 
For $d_{xy}$ OP the reversed power laws for $B_{1g}$ and $B_{2g}$ 
channels are expected.
\end{thebibliography}
\end{document}